\documentclass[%
aps,prd,twocolumn,
superscriptaddress,
]{revtex4-2}

\usepackage{graphicx}
\usepackage{amsfonts}
\usepackage{amssymb}
\usepackage{amsthm}
\usepackage{amsmath}
\usepackage{multirow}

\usepackage{color}
 \usepackage{slashed}
\def\beq{\begin{equation}}
\def\eeq{\end{equation}}
\def\bea{\begin{eqnarray}}
\def\eea{\end{eqnarray}}
\newcommand{\beqs}{\begin{subequations}}
\newcommand{\eeqs}{\end{subequations}}

\newcommand{\cref}[1]{Ref.~\cite{#1}}

\newcommand{\hh}{{\ensuremath{I{\kern-2.6pt h}}}}
\newcommand{\bhh}{{\ensuremath{\bar{I{\kern-2.6pt h}}}}}

\usepackage[colorlinks=true,allcolors=blue]{hyperref}
\usepackage{mathtools}

\usepackage[thinc]{esdiff}

\begin{document}
	
\title{C-parity, magnetic monopoles and higher frequency gravitational waves}
\author{Rinku Maji}
\affiliation{Cosmology, Gravity and Astroparticle Physics Group, Center for Theoretical Physics of the Universe,  Institute for Basic Science, Daejeon 34126, Republic of Korea}
	\author{Qaisar Shafi}
		\affiliation{Bartol Research Institute, Department of Physics and 
		Astronomy,
		 University of Delaware, Newark, DE 19716, USA}
\begin{abstract}
We consider the spontaneous breaking of $SO(10)$ grand unified symmetry to the left-right symmetric model $SU(3)_c \times SU(3)_L \times SU(2)_R \times U(1)_{B-L}$ with C-parity also unbroken [$C$ converts $Q\to -Q$, where $Q$ is the electric charge operator in $SO(10)$.] This breaking produces the topologically stable GUT monopole as well as a GUT scale C-string. The subsequent breaking at an intermediate scale of C-parity produces domain walls bounded by C-strings, found by Kibble, Lazarides and Shafi. A limited number of inflationary $e$-foldings experienced during these breakings can yield an observable number density of primordial GUT monopoles. The C-strings also experience this inflationary phase, and the subsequent string-wall network decays through the emission of gravitational waves. We estimate the gravitational wave spectrum from these composite structures over a range of values of the domain wall tension
$\sigma$. Depending on $\sigma$ the spectrum displays a peak in the higher frequency range 
between $10^2$ to $10^5$ Hz.
\end{abstract}
\maketitle

\section{Introduction}
As a grand unified gauge symmetry $SO(10)$ \cite{Georgi:1974my, Fritzsch:1974nn} has several key features. Its 16 dimensional spinor representation contains the matter fields of the Standard Model (SM) single family as well as a right handed neutrino. One of its maximal subgroups is $SU(4)_c \times SU(2)_L \times SU(2)_R$ \cite{Pati:1974yy} which displays left-right symmetry and lepton number is the fourth color.
In addition, $SO(10)$ contains two important discrete gauge symmetries. One of them is the $Z_2$ subgroup of $Z_4$ \cite{Kibble:1982ae}, the centre of $SO(10)$. Breaking $SO(10)$ to the SM and subsequently to $SU(3)_c \times U(1)_{\rm em}$ with all Higgs scalars in tensor representations leaves this $Z_2$ symmetry unbroken, which yields topologically stable cosmic strings \cite{Kibble:1982ae}. In a supersymmetric setting this $Z_2$ symmetry is precisely ``matter parity", and provides a key ingredient for a stable dark matter particle. For a recent discussion of dark matter with this $Z_2$ symmetry in a non-supersymmetric $E_6$ model, see Ref.~\cite{Maji:2024pll}. A second important discrete symmetry in $SO(10)$ is C-parity \cite{Kibble:1982ae,Kibble:1982dd,LAZARIDES1985261}, also known as D-parity \cite{Chang:1983fu}, which interchanges left and right chiral fields accompanied with complex conjugation. Acting on the electric charge generator it changes $Q$ to $-Q$ \cite{Kibble:1982dd,LAZARIDES1985261}. More explicitly, following \cite{LAZARIDES1985261}, $CQC^{-1}=-Q$, where $C=(i\sigma_{23})(i\sigma_{67})(i\sigma_{89})$ and $\sigma_{ij}$ ($i=1,2,...,10$) are the generators of $SO(10)$.

It was shown in \cite{Kibble:1982ae,Kibble:1982dd} that the breaking of $SO(10)$ to a left-right symmetric model $H$ that also leaves $C$ unbroken yields a C-string. This is because the unbroken symmetry group  in this case has two disconnected components, namely $H$ and $CH$. However, the C-string is not topologically stable because C-parity is spontaneously broken together with $H$. This yields the well known composite structure dubbed ``wall bounded by string" (WBS) \cite{Kibble:1982dd}, which were recently discovered in superfluid $^3$He-B \cite{Makinen:2018ltj}. These WBS structures are now known to also arise in Nematic liquid crystals and in spin systems \cite{Takeuchi:2020ptu}.

In this paper we explore the cosmological imprints of C-parity. Note that the C-string is associated with the GUT scale and is therefore superheavy. The breaking scale of $C$ determines the domain wall tension, and the interplay between these two scales as well as the role of cosmic inflation will determine the cosmological signatures that we wish to explore. It is important to note that the breaking of $SO(10)$ to a left-right symmetric model $H$ also produces the superheavy GUT monopole. In other words, the $C$ string is accompanied by the superheavy monopole, and inflation is employed here to suppress the monopole number density to acceptable levels. This, of course, also impacts the gravitational wave spectrum emitted by the string-wall system as we shall see. Ref.~\cite{Lazarides:2024niy} shows that the decay of metastable cosmic strings can produce higher frequency gravitational waves along with an observable flux of monopoles. For recent studies on string-wall hybrid structures see Refs.~\cite{Maji:2023fba,Lazarides:2023ksx,Ge:2023rce, Eto:2023aqr, Hamada:2024dan, Ringe:2024ktt, Fujikura:2024biv, Bao:2024bws, Roshan:2024qnv}.

The paper is laid out as follows. In Section \ref{sec:2} we display a variety of $SO(10)$ breaking schemes via the left-right symmetric groups $SU(4)_c \times SU(2)_L \times SU(2)_R$ and $SU(3)_c \times SU(3)_L \times SU(2)_R \times U(1)_{B-L}$, which are accompanied by an unbroken C-parity. For completeness, we show that the breaking of $SO(10)$ to $SU(3)_c \times SU(2)_L \times SU(2)_R\times U(1)_{B-L}$ yields the topologically stable superheavy GUT monopole. [For the breaking of $SO(10) \to SU(4)_c \times SU(2)_L \times SU(2)_R$, the proof can be found in Ref.~\cite{Lazarides:2019xai} and references therein.] In Section \ref{sec:dyn-times} we address the important challenge of reducing the GUT monopole density to acceptable, indeed observable levels. This, we assume, can be realized in suitable inflationary models where the monopoles experience a controlled number of $e$-foldings. For recent discussion of how this is achieved in hybrid inflation models, see Refs.~\cite{Lazarides:2023rqf,Moursy:2024hll,Maji:2024cwv}.
This, of course, has implications for the C-strings as well as the string-wall system that emerges from the spontaneous breaking of C-parity at an intermediate scale. We discuss in this section the timescales for the domain wall formation, string dynamics domination, and a domain wall dominated universe. In Section \ref{sec:GWs-WBS} we discuss the gravitational wave spectrum emitted by the string-wall network and show that the spectrum primarily lies in the higher frequencies with a peak around $10^2-10^5$ Hz.
Our conclusions are summarized in Section \ref{sec:conc}.

\section{$SO(10)$ symmetry breaking, C-parity and topological defects}
\label{sec:2}

\begin{figure}[h!]
\begin{center}
\includegraphics[width=0.99\linewidth]{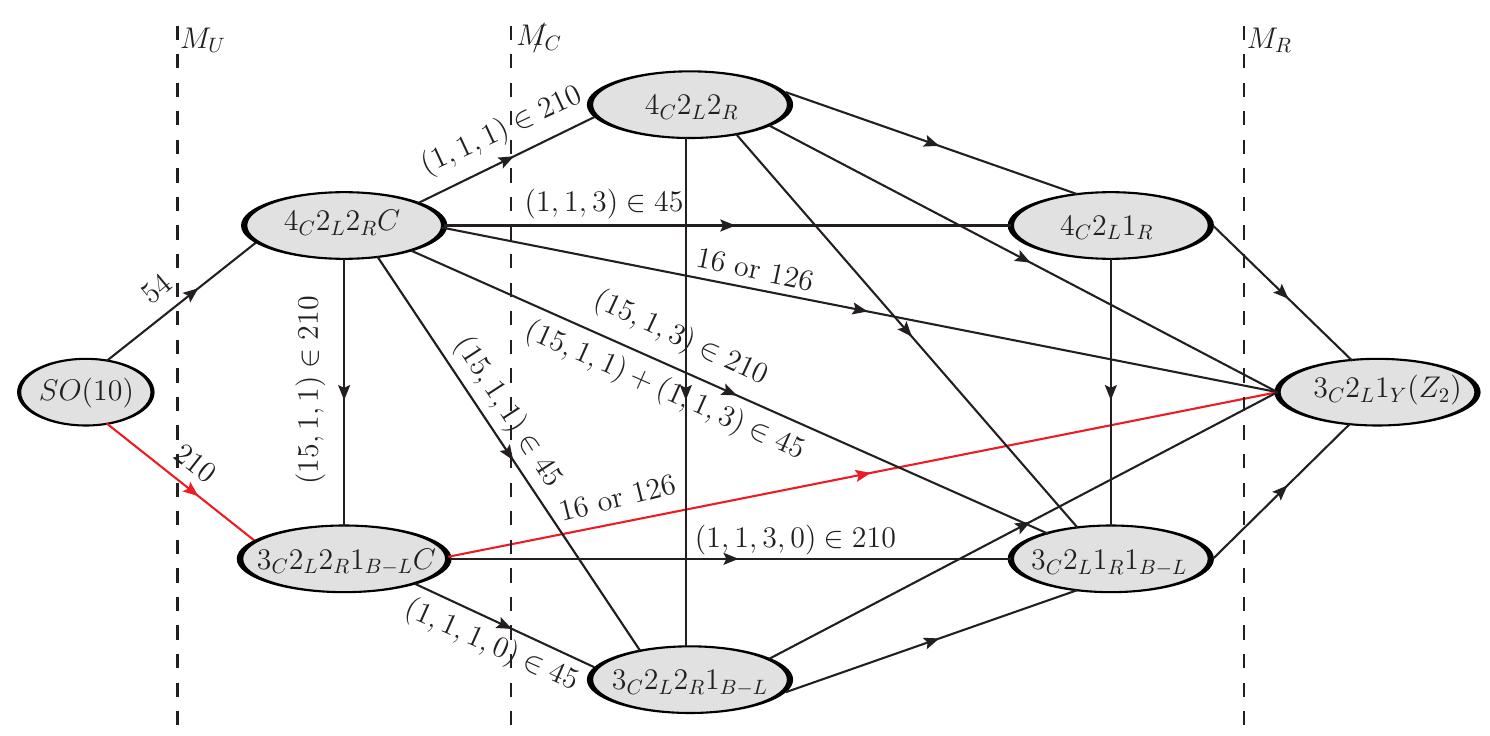}
\end{center}
\caption{$SO(10)$ breaking at GUT scale with C-parity unbroken proceeds via the two left-right gauge symmetries shown. For completeness, we display a variety of subsequent symmetry breaking chains. In this paper we focus on the simplest breaking chain $SO(10) \to SU(3)_c \times SU(2)_L \times SU(2)_R \times U(1)_{B-L} \to SU(3)_c \times SU(2)_L \times U(1)_Y$, which is highlighted in red.
}
\label{fig:breaking-chains}
\end{figure}

The breaking of $SO(10)$, more precisely Spin(10), to a left-right symmetric group such as $SU(4)_c \times SU(2)_L \times SU(2)_R$ ($4_C2_L2_R$, for short), or its subgroup $SU(3)_c \times SU(2)_L \times SU(2)_R \times U(1)_{B-L}$ ($3_C2_L2_R1_{B-L}$), produces a superheavy topologically stable magnetic monopole. This is because the second homotopy group of $G/H$ in this case is non-trivial, with $G = SO(10)$ and $H = 4_C2_L2_R$ or $3_C2_L2_R1_{B-L}$. It was shown in Ref.~\cite{Kibble:1982ae} that for a suitable choice of Higgs scalar and it's VEV, the first homotopy group of
$G/H$ can also be non-trivial, which implies the existence of cosmic strings. In particular, if C-parity is unbroken at $M_{\rm GUT}$, we expect the appearance of C-strings  (see Fig.~\ref{fig:breaking-chains} for the breaking patterns). However, since C-parity is broken in nature, the C-strings end up as boundaries of domain walls, as shown in Ref.~\cite{Kibble:1982dd}.
It is important to recall that the breaking of $4_C2_L2_R$ produces a new set of topologically stable magnetic monopoles as described in Refs.~\cite{Lazarides:1980cc,Lazarides:2019xai}. This, however, is not the case if the $SO(10)$ breaking proceeds via $3_C2_L2_R1_{B-L}$, which simplifies our investigation of the composite string-wall structures and emission of gravitational waves. The strings and walls respectively have a tension $\mu = \pi v_U^2$ and $\sigma = \frac{2\sqrt{2}}{3} \sqrt{\lambda}v_{\rm dw}^3$, where $v_i$ ($i=U,{\rm dw}$) denotes the VEVs of scalars associated with the unification and wall forming symmetry breaking. The parameter $\lambda$ represents the effective value of the scalar quartic coupling which we vary from $0.1$ to 10. It depends on the details of the scalar potential (see, for example, Ref.~\cite{Pogosian:2000xv}).
The simultaneous appearance of both superheavy and intermediate mass topologically stable magnetic monopoles in the breaking via $4_C2_L2_R$ requires a separate investigation.

For completeness, let us show that the breaking of $SO(10)$ to a left-right symmetric model $H$ produces the topologically stable superheavy GUT monopole. We set $H = SU(3)_c \times SU(2)_L \times SU(2)_R \times U(1)_{B-L}$ since it appears in the breaking chain we focus on here. To identify the GUT monopole we consider the first homotopy group of $H$ and show that it is non-trivial.
As topological spaces the compact group $U(1)_{B-L}$ and $SU(3)_c$ intersect in the $Z_3$ center of the latter. Consider a $2 \pi$ rotation with $\frac{2}{3} Y_c^8$, where $Y_c^8$ (= diag($1,1,-2$)) denotes the color hypercharge generator. Acting on the color triplets it brings us to the center of $SU(3)_c$, with the lepton fields remaining unchanged.

Next we can perform a $\pi$ rotation along $B-L$ which, in combination with the previous rotation, means that this effectively is a $Z_2$ rotation on the quark and lepton fields thereby changing their sign. Since $H$ is embedded inside $SO(10)$, the $Z_2$ subgroups in $SU(2)_L$ and $SU(2)_R$
intersect the $Z_2$ transformation generated by $\frac{2}{3} Y_c^8 + (B-L)$, as shown in \cite{Lazarides:1980cc,Lazarides:2019xai}. Noting that the leptons and quarks reside in the doublet representations of $SU(2)_L$ and $SU(2)_R$, we can perform rotations by $\pi$ with the generators $T_L^3$ and $T_R^3$ on these doublets, which finally brings us to the identity element.

In summary, we have shown the existence of a non-trivial loop in $H$ which is generated by a $2 \pi$ rotation along $Q + \frac{1}{3} Y_c^8$, where the electric charge generator $Q = T_L^3/2 + T_R^3/2 + (B-L)/2$. Thus, we have identified the $SO(10)$ GUT monopole which carries a single unit ($2 \pi/e$) of Dirac magnetic charge as well as color magnetic charge.

Before closing this section, it is worth mentioning that a second and distinct set of ``walls bounded by strings" structures can be realized in this $SO(10)$ model. The string in this case appears from breaking of $U(1)_{B-L}$ to $Z_2$ with the 126-dimensional representation, and therefore it’s mass per unit length is associated with the seesaw scale \cite{Kibble:1982ae}. A subsequent breaking of $Z_2$ with a scalar in the 16-dimensional representation means that the string forms the boundary of a domain wall \cite{Lazarides:2023iim, Maji:2023fba}. We will not discuss the cosmological implications of this structure in the paper.

\section{Monopole flux, string-wall dynamics and timescales}
\label{sec:dyn-times}
The extended structures ``walls bounded by strings" can decay after a time scale $R_c=\mu/\sigma$ as the string dynamics start dominating in a WBS. Since the strings are produced along with topologically stable monopoles at the GUT scale, they should experience some $e$-foldings so that the monopole flux lies below the experimental bounds \cite{Ambrosio:2002qq,IceCube:2021eye} (Table~\ref{tab:exp-lim}).
\begin{table}[htbp!]
\begin{center}
\begin{tabular}{|c| c| c| c|}
\hline
\multirow{2}{*}{Experiment} & {Monopole mass} & {Average } & Flux $\mathcal{F}_{M}$ in \\
& ($m_M$) in GeV &velocity ($v_{M}$)&  $\mathrm{cm}^{-2}\mathrm{sec}^{-1}\mathrm{sr}^{-1}$ \\ 
\hline
{MACRO \cite{Ambrosio:2002qq}} 
& $>5\times 10^{13}$  & $>4\times 10^{-5}$ & $1.4\times 10^{-16}$ \\
\hline
{IceCube \cite{IceCube:2021eye}} & $>10^{8}-10^{10}$ & $0.8 - 0.995$ & $2\times 10^{-19}$ \\
\hline
\end{tabular}
\caption{Upper limits on the monopole flux from MACRO and IceCube \cite{Ambrosio:2002qq,IceCube:2021eye, Patrizii:2015uea}.}\label{tab:exp-lim}
\end{center}
\end{table}
 We can re-express the bound on the monopole flux as a bound on the monopole yield given by
\begin{align}\label{eq:mon-yield}
Y_M\equiv &\frac{n_{M}}{s}=\frac{4\pi \mathcal{F}_M}{v_M s_0} \nonumber\\ \simeq & 2\times 10^{-10}\left(\frac{\mathcal{F}_M }{\mathrm{cm}^{-2}\mathrm{sec}^{-1}\mathrm{sr}^{-1}}\right)\left(\frac{v_M}{10^{-3}}\right)
\end{align}
where $s_0$ denotes entropy density, $v_M$ is the average velocity of the  monopoles, and $\mathcal{F}_M$ denotes the present day flux of monopoles.

\begin{figure}[h!]
\begin{center}
\includegraphics[width=0.99\linewidth]{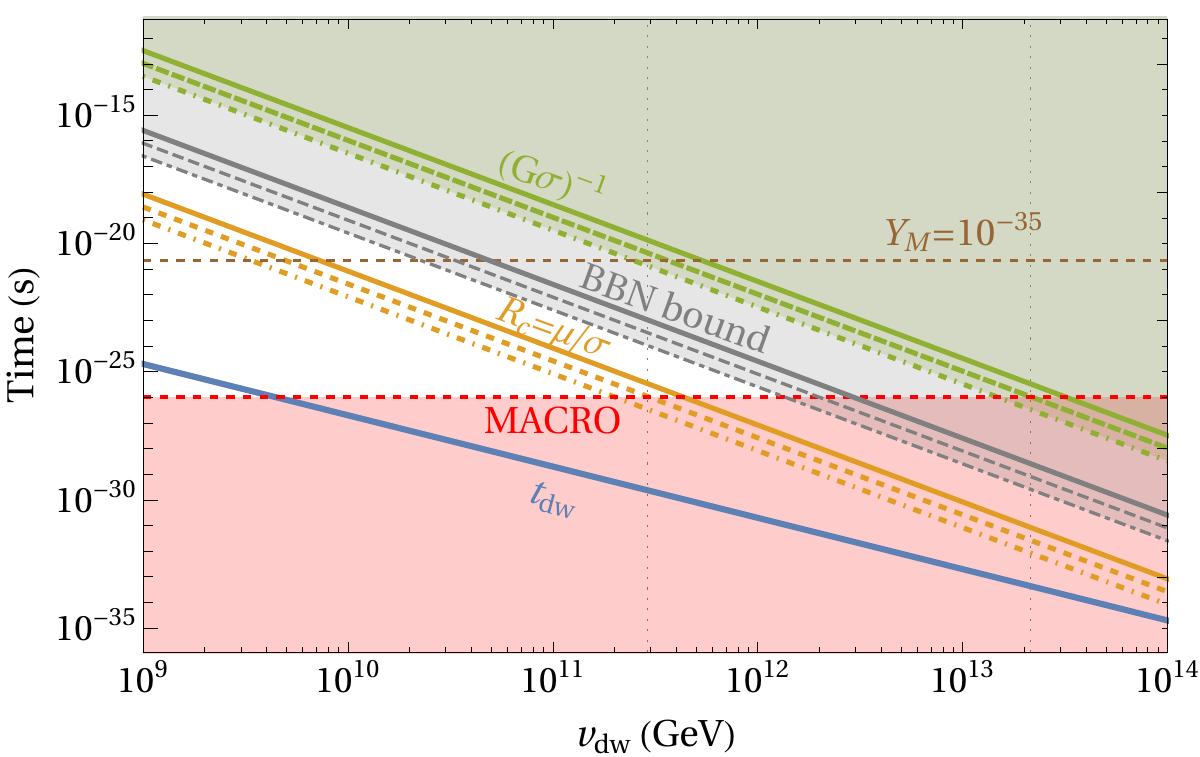}
\end{center}
\caption{Timescales for domain wall formation, string dynamics domination, and domain wall dominated universe for a GUT scale $v_U=10^{16}$ GeV. We have shown in blue the formation time of the C-walls in the radiation dominated universe. The radius of curvature of WBS (which is of the order of the cosmic time), $R_c$, after which the string dynamics dominate over the walls is plotted in orange. The domain walls dominate the energy budget of the universe after a timescale $(G\sigma)^{-1}$, which is shown by the green line. Coincidentally, this is also the timescale after which a WBS inside the particle horizon collapse. The green shaded region shows the timescale when the string cannot re-enter the horizon in a wall dominated universe. The red dashed line arises from Eqs.~\eqref{eq:mon-yield} and \eqref{eq:mon-yield-time} and shows the MACRO bound on the monopole flux with average velocity $10^{-3}$. The horizon re-entry of strings and monopoles before this timescale is, therefore, excluded as dictated by the red shaded region. The brown dashed line is an adopted threshold for observability. The gray shaded region is excluded by the Big Bang Nucleosynthesis (BBN) bound on any additional effective number of relativistic degrees of freedom  $\Delta N_{\rm eff}$ beyond the Standard Model neutrinos (see Sec.~\ref{sec:GWs-WBS} for more details). The solid, dashed and dotted-dashed lines in the plots of $(G\sigma)^{-1}$, $R_c$ and BBN bound represent the timescales for three choices of $\lambda=0.1$, $1$ and $10$ respectively.}
\label{fig:timescale}
\end{figure}
Now suppose the monopoles and strings suffer partial inflation and re-enter the horizon at time $t_F$ before or after the reheat time $t_r$, which can be  estimated from the time-temperature relation during radiation domination:
\begin{align}\label{eq:t-T}
T^2 = \sqrt{\frac{45}{2\pi^2g_*}}\frac{m_{\rm Pl}}{t},
\end{align}
where $g_*$ is the effective number of relativistic degrees of freedom contributing to the energy density and $m_{\rm Pl}=1/\sqrt{8\pi G}$ ($G$ is the Newton's constant) is the reduced Planck mass.
The entropy density at time $t$ is expressed as
\begin{align}
s(t)=\frac{2\pi^2}{45}g_{*s} T(t)^3,
\end{align}
with $g_{*s}$ being the effective number of relativistic degrees of freedom for the entropy density. There will be order one monopole within a Hubble volume, $V_h = (2t_F)^3$, at the horizon re-entry time $t_F$ of the C-strings as they are formed together with the GUT monopoles.
 We estimate the yield of monopoles experiencing partial inflation to be
\begin{align}
\label{eq:mon-yield-time}
Y_M\simeq\frac{1}{V_h s(t_F)}\simeq 
\begin{cases}
10^{-65}\frac{g_*(t_F)^{3/4}}{g_{*s}(t_F)}\left(\frac{\mathrm{sec}}{t_F}\right)^{3/2} & \mathrm{for} \ \ t_F\geq t_r , \\
10^{-65}\frac{g_*(t_r)^{3/4}}{g_{*s}(t_r)}\frac{\mathrm{sec}^{3/2}}{t_Ft_r^{1/2}} & \mathrm{for} \ \ t_F\leq t_r.
\end{cases}
\end{align}
The lower bound on the horizon re-entry time of the monopoles and C-strings is, therefore, $t_F\gtrsim 10^{-26}$ seconds. The formation time ($t_{\rm dw}$) of the walls can be obtained from Eq.~\eqref{eq:t-T} with $T=v_{\rm dw}$. If the superheavy monopoles along with the C-strings are completely inflated away, or if the GUT symmetry never restored \cite{Weinberg:1974hy, Mohapatra:1979qt, Mohapatra:1979vr, Mohapatra:1979bt, Dvali:1995cc, Dvali:1995cj}, the domain walls will dominate the universe after a timescale $(G\sigma)^{-1}$  which, of course, is unacceptable. 
 
 Fig.~\ref{fig:timescale} shows different timescales as a function of the domain wall vev $v_{\rm dw}$. The monopoles and C-strings should re-enter the horizon within the timescale consistent with the MACRO bound (red dashed line) and  $(G\sigma)^{-1}$. The WBS structures decay by radiating gravitational waves after $t_F$  which we discuss in the next section.
\section{Gravitational waves from walls bounded by strings}
\label{sec:GWs-WBS}
The domain walls are stable before the horizon re-entry of the strings. These WBS structures after $t_F$ are dominated by the string dynamics until the time $R_c$  if $t_F<R_c$.  The strings can intercommute, form loops, and enter a scaling regime at a later time $t_s$, if $t_s<R_c$ \cite{Vachaspati:1984gt,Kibble:1984hp,Vilenkin:2000jqa,Damour:2001bk,Vanchurin:2005pa,Ringeval:2005kr,Olum:2006ix,Leblond:2009fq, Olmez:2010bi,Blanco-Pillado:2013qja,Blanco-Pillado:2017oxo,Cui:2018rwi,Buchmuller:2019gfy,Buchmuller:2021mbb,Dunsky:2021tih}. On the other hand, the wall dynamics dominates for $t_F\gtrsim R_c$. The gravitational wave spectrum will be dominated by the radiation emitted by the string loops bounding the walls. The gravitational waves in the later case will arise from the oscillating and collapsing strings connected by the walls structures.
\begin{figure}[h!]
\begin{center}
\includegraphics[width=0.99\linewidth]{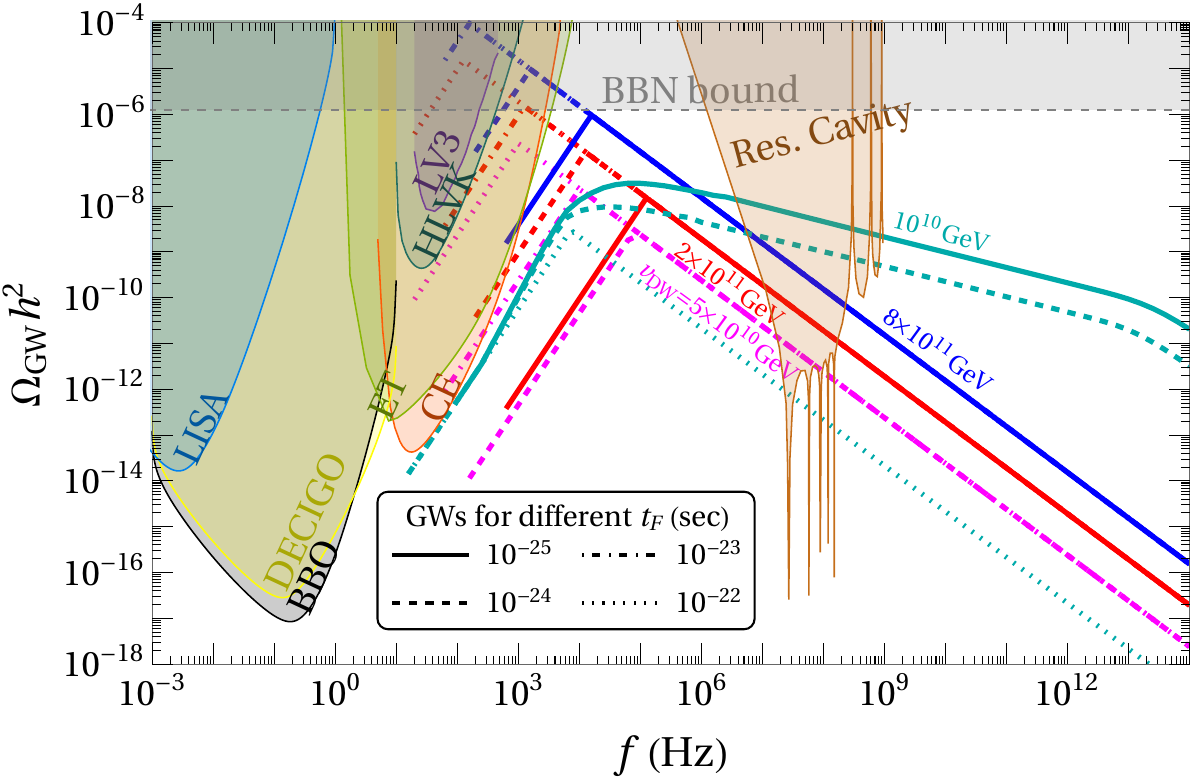}
\end{center}
\caption{Stochastic gravitational wave background from WBS structures for the benchmark choices of the intermediate scale VEV, $v_{\rm dw} = \lbrace 10^{10}, 5\times 10^{10}, 2\times 10^{11}, 8\times 10^{11}\rbrace$ GeV, associated with the breaking of C-parity, for different choices of $t_F=10^{-25}-10^{-22}$ seconds and $v_U=10^{16}$ GeV. For $v_{\rm dw}=10^{10}$ GeV and $t_F\sim 10^{-25}-10^{-24}$ sec, the gravitational wave spectra (cyan) is dominated by the radiation from the strings. For larger $t_F>10^{-23}$ sec, the string enter the scaling regime or re-enter the horizon at or after $R_c$. The gravitational waves in these cases are generated from the oscillating and collapsing WBS similar to the choices of $v_{\rm dw}$  $ 5\times 10^{10}$ GeV (magenta), $2\times 10^{11}$  GeV (red), and $8\times 10^{11}$ GeV (blue).  We have depicted the power law integrated sensitivity curves \cite{Thrane:2013oya, Schmitz:2020syl} for planned experiments in the mHz to few kHz frequencies such as DECIGO \cite{Sato_2017}, BBO \cite{Crowder:2005nr, Corbin:2005ny}, LISA \cite{Bartolo:2016ami, amaroseoane2017laser}, HLVK \cite{KAGRA:2013rdx}, ET \cite{Mentasti:2020yyd}, and  CE \cite{Regimbau:2016ike}. A sensitivity curve of resonant electromagnetic cavities claimed in Refs.~\cite{Herman:2020wao,Herman:2022fau} is shown in the brown shaded region which is challenged in a recent review \cite{Aggarwal:2025noe}. The gray dashed line depicts the BBN bound on the gravitational wave background.  We set $\lambda = 1$ to estimate the gravitational waves.}
\label{fig:GWs}
\end{figure}
\subsection{String dynamics domination before timescale $R_c$}
\label{subsec:GWs-I}
 The horizon re-entry time $t_F$ of the strings  in our case is much later than the time scale of early friction domination, $t_{\rm Pl}/(G\mu)^2$ \cite{Vilenkin:1991zk, Garriga:1993gj}, where $t_{\rm Pl}$ is the Planck time and domination of particle emission \cite{Blanco-Pillado:1998tyu,Matsunami:2019fss, Auclair:2019jip}. Before the timescale $R_c$ when string dynamics dominates, the string network enters the scaling regime at a time $t_s$, which is typically two orders of magnitude higher than $t_F$ \cite{Martins:1995tg, Martins:1996jp, Martins:2000cs,Gouttenoire:2019kij}. 
The gravitational wave background can be expressed as
\begin{align}
\Omega_{\rm GW}(f) = \sum_{k=1}^{\infty}\Omega_{\rm GW}^{(k)}(f), \quad k\in Z^+,
\end{align}
where the gravitational waves from a normal mode is given by \cite{Dunsky:2021tih,Maji:2023fba}
\begin{align}
   \label{eq:GWs2}
    \Omega^{(k)}_{\rm GW}(f) 
    = &\frac{1}{\rho_{c,0}} \int_{t_s}^{t_0} d\tilde{t} \left(\frac{a(\tilde{t})}{a(t_0)}\right)^5\frac{\mathcal{F} C_{\rm eff}(t_i)}{(\Gamma G \mu + \alpha)\alpha t_i^4} \left(\frac{a(t_i)}{a(\tilde{t})}\right)^3 \nonumber \\ & \frac{\Gamma k^{-4/3}}{\zeta(4/3)} G\mu^2 \frac{2 k}{f}\Theta(R_c - t_i) .
\end{align}
Here $a(t)$ denotes the scale factor, $\rho_{c,0}$ is the present day ($t_0$) critical energy density of the universe, $\mathcal{F}\simeq 0.1$, $\Gamma \simeq 50$, $\alpha\simeq 0.1$ for string network in the scaling regime \cite{Vachaspati:1984gt,Vilenkin:2000jqa}, and  $C_{\rm eff} = 5.7$ in the radiation dominated era \cite{Vanchurin:2005pa,Ringeval:2005kr,Olum:2006ix,Olmez:2010bi,Blanco-Pillado:2013qja,Blanco-Pillado:2017oxo,Cui:2018rwi}. 

\subsection{Wall dynamics domination after timescale $R_c$}
\label{subsec:GWs-II}
This section discusses the gravitational waves from the oscillating and collapsing WBS after time $R_c$. The WBS structures oscillate with constant physical size before collapsing at a time scale $t_d\sim 1/(G\sigma)$. The physical size of an oscillating WBS is $w \sim 2t_F$, and the power radiated from a WBS structure at a time $t$ is given by \cite{Hiramatsu:2013qaa,Dunsky:2021tih,Bao:2024bws}
\begin{align}
P_{\rm GW} = \diff{E_{\rm GW}}{t}\approx G\sigma^2 w l ,
\end{align}
where $l = 2t$ is the Hubble radius at the time $t$. The WBS structures oscillate with constant size before collapsing at $t_d$. The gravitational wave background at $t_d$ from the oscillating WBS is given by
\begin{align}\label{eq:GWs-WBS-osc}
\Omega_{\rm GW}^{\rm osc}(\tilde{f},t_d) = \frac{\sigma w}{m_{\rm Pl}^2} (\tilde{f} w)^2 \quad \mathrm{with} \ \frac{1}{w}\frac{a(t_F)}{a(t_d)}\leq \tilde{f} \leq \frac{1}{w}
\end{align}
where $\tilde{f}=(1+z_d)f$, with $f$ being the redshifted frequency at present epoch. At time $t_d$, the WBS structures collapse and produce the ultraviolet tail of the gravitational wave spectra given by
\begin{align}\label{eq:GWs-WBS-col}
\Omega_{\rm GW}^{\rm col}(\tilde{f},t_d) = \frac{\sigma w}{m_{\rm Pl}^2} (\tilde{f} w)^{-1}, \quad \mathrm{with} \  \tilde{f} \geq \frac{1}{w} \ .
\end{align}
Finally, the gravitational wave spectrum is given by
\begin{align}
\Omega_{\rm GW}(f)=\mathcal{G}(z_d)\Omega_{r,0}\Omega_{\rm GW}(\tilde{f},t_d),
\end{align}
where $\Omega_{r,0}=9.1476\times 10^{-5}$ \cite{Planck:2018vyg} is the present fractional energy density of radiation, and \cite{Binetruy:2012ze}
\begin{align}
\mathcal{G}(z)=\frac{g_*(z)}{g_*(0)}\frac{g_{*s}(0)^{4/3}}{g_{*s}(z)^{4/3}} .
\end{align}

Fig.~\ref{fig:GWs} depicts the gravitational waves from the WBS structures for the benchmark choices of the VEV, $v_{\rm dw} = \lbrace 10^{10}, 5\times 10^{10}, 2\times 10^{11}, 8\times 10^{11}\rbrace$ GeV, associated with the breaking of C-parity, and $v_U=10^{16}$ GeV. We have shown the spectra for $t_F$ varying from $10^{-25}-10^{-22}$ sec. For $v_{\rm dw}=10^{10}$ GeV, the gravitational wave background is dominated by the radiation from the strings for $t_F\sim 10^{-25}-10^{-24}$ sec. (see Fig.~\ref{fig:timescale}). For $t_F>10^{-23}$ sec, the strings enter the scaling regime or re-enter the horizon at or after $R_c$. The gravitational waves in this case will arise from the oscillating and collapsing WBS similar to the cases when $v_{\rm dw} = \lbrace 5\times 10^{10}, 2\times 10^{11}, 8\times 10^{11}\rbrace$ GeV.  The next generation of proposed ground based detectors, namely, HLVK \cite{KAGRA:2013rdx}, ET \cite{Mentasti:2020yyd}, and CE \cite{Regimbau:2016ike} will be sensitive for $v_{\rm dw}\sim 10^{11}$ GeV with the horizon re-entry time $t_F\gtrsim 10^{-24}$ sec, as can be seen from the power-law integrated sensitivity curves  \cite{Thrane:2013oya, Schmitz:2020syl}.  High frequency gravitational wave detectors such as resonant electromagnetic cavities may be sensitive to the high frequency ultraviolet tails of the gravitational wave spectra as discussed in Refs.~\cite{Herman:2020wao,Herman:2022fau}. However, this claim has been contradicted in a recent review \cite{Aggarwal:2025noe}.

The gravitational waves are radiated before big bang nucleosynthesis (BBN) and cosmic microwave background (CMB) decoupling, and therefore constrained by the measurement of the effective number of `neutrino' degrees of freedom $N_{\rm eff}$ \cite{Aver:2015iza, Peimbert:2016bdg, Planck:2018vyg}. The effective number of neutrinos in the SM is $N_{\rm eff}^{\rm SM}=3.044(1)$ \cite{EscuderoAbenza:2020cmq,Akita:2020szl,Froustey:2020mcq,Bennett:2020zkv}. The combined upper bound from BBN and CMB  on any additional effective number of relativistic degrees of freedom is $\Delta N_{\rm eff}\equiv N_{\rm eff}-N_{\rm eff}^{\rm SM}\lesssim 0.22$  \cite{Planck:2018vyg}.
 The constraint on the gravitational radiation reads
\begin{equation}\label{eq:bbn-cmb}
\int\Omega_{\rm GW}(f)d\ln{f} \lesssim \Omega_{\gamma,0} \frac{7}{8}\left(\frac{4}{11}\right)^{4/3}\Delta N_{\rm eff} ,
\end{equation}
where $\Omega_{\gamma,0}\simeq 2.5\times 10^{-5}$ is the present day energy fraction in photons. In Fig.~\ref{fig:GWs}, we have depicted the BBN bound on the gravitational wave spectrum emitted from the oscillating and collapsing WBS structures. In fact, the bound comes from the peak of the spectra and depends on the wall tension $\sigma$ and horizon re-entry time $t_F$ (see Eqs.~\eqref{eq:GWs-WBS-osc} and \eqref{eq:GWs-WBS-col}), which has been displayed in Fig.~\ref{fig:timescale}.

\section{Conclusions}
\label{sec:conc}
 It is widely discussed that the very early universe may have experienced a variety of phase transitions with potentially far reaching consequences.
Motivated by the discovery of gravitational waves, in this paper we have explored if the phase transitions associated with a well motivated $SO(10)$ model can leave any observable cosmological imprints in the present universe. The $SO(10)$ symmetry breaking in our scenario proceeds via a left-right symmetric model that also leaves unbroken the discrete C-parity (which changes the electric charge generator $Q$ to $-Q$.) This phase transition produces the superheavy GUT magnetic monopoles as well as C-strings. In the next breaking which takes place at an intermediate scale, the C-parity is spontaneously broken such that the strings become boundaries of domain walls, as first shown in Ref.~\cite{Kibble:1982dd}.
We require that cosmic inflation dilutes the number density of the GUT magnetic monopoles to observable levels, chosen to be a few orders of magnitude below the MACRO bound. The C-strings experience the same amount of inflation as the monopoles, which, in turn, impacts the gravitational wave spectrum emitted by the composite string-domain wall system. We find that compatibility with the CMB measurements and other observations yields a spectrum in a higher frequency range with its peak varying from $10^2 -10^5$ Hz. Thus, an observable number density of superheavy GUT monopoles and a novel gravitational wave spectrum emitted by the string-wall network are two testable predictions arising from the phase transitions associated with the $SO(10)$ breaking in the very early universe.

\section{Acknowledgment}
RM is supported by the Institute for Basic Science under the project code: IBS-R018-D3.

\bibliographystyle{JHEP}
\bibliography{Cparity_cleaned}

\end{document}